\documentclass[11pt]{article}
\usepackage{fullpage,amssymb,url,amsmath,amsthm}
\usepackage{graphicx}
\usepackage{authblk}

\def\der{\mathrm{d}}
\def\dmu{\mathrm{d}\mu}
\def\dx{\mathrm{d}x}
\def\dy{\mathrm{d}y}
\def\dt{\mathrm{d}t}
\def\KL{\mathrm{KL}}

\def\bbQ{{\mathbb{Q}}}
\def\bbR{{\mathbb{R}}}

\def\calL{\mathcal{L}}

\newtheorem{theorem}{Theorem}

\newtheorem{proposition}{Proposition}

\title{A closed-form formula for the Kullback-Leibler divergence between Cauchy distributions}

\author[$\dagger$]{Fr\'ed\'eric Chyzak}
\affil[$\dagger$]{INRIA}
\affil[$\dagger$]{Universit\'e Paris--Saclay, France}
\affil[$\dagger$]{{\small\tt frederic.chyzak@inria.fr}}
\author[$\star$]{Frank Nielsen}
\affil[$\star$]{Sony Computer Science Laboratories, Inc.}
\affil[$\star$]{Tokyo, Japan}
\affil[$\star$]{{\small\tt Frank.Nielsen@acm.org}}

\date{}

\begin{document}
\maketitle

\begin{abstract}
We report a closed-form expression for the Kullback-Leibler divergence between Cauchy distributions which involves the calculation of a parametric definite integral with 6~parameters. The formula shows that the Kullback-Leibler divergence between Cauchy densities is always finite and symmetric.
\end{abstract}

\section{Introduction}

The Cauchy distribution $P_{l,s}$ has the following density, $p_{l,s}$, with respect to the Lebesgue measure~$\mu$ on $(-\infty,\infty)$:

\begin{equation}
p_{l,s}(x) = \frac{\mathrm{d}P_{l,s}}{\mathrm{d}\mu}(x) = \frac{s}{\pi (s^2+(x-l)^2)},
\end{equation}
where $l$ and $s$ denote the location parameter and scale parameter, respectively.

The set of Cauchy densities form a location-scale family~\cite{Murray-1993,KLls-2019} with Cauchy standard distribution
\begin{equation}
p(x)=p_{0,1}(x)=\frac{1}{\pi (1+x^2)}.
\end{equation}

Indeed, for any density $p$ (called the standard density), one can generate a location-scale family $\{p_{l,s} \ |\ l\in\bbR, s>0\}$ by defining the density
\begin{equation}
p_{l,s}(x)=\frac{1}{s}p\left(\frac{x-l}{s}\right).
\end{equation}

We want to calculate the Kullback-Leibler (KL) divergence~\cite{CT-2012} between two Cauchy densities $p_{l_1,s_1}$ and $p_{l_2,s_2}$:
\begin{equation}
\KL(p_{l_1,s_1}:p_{l_2,s_2})=\int p_{l_1,s_1}(x)\log \frac{p_{l_1,s_1}}{p_{l_2,s_2}} \dmu(x). \label{eq:KLls}
\end{equation}
Since the KL divergence is the relative entropy~\cite{CT-2012} (i.e., the difference between the cross-entropy and the entropy), we have the following identity:
\begin{equation} \label{eq:KL-from-h}
\KL(p_{l_1,s_1}:p_{l_2,s_2})= h^\times(p_{l_1,s_1}:p_{l_2,s_2})-h(p_{l_1,s_1}),
\end{equation}
where $h^\times$ denotes the cross-entropy and $h(p_{l_1,s_1})=h^\times(p_{l_1,s_1}:p_{l_1,s_1})$ the differential entropy~\cite{h-handbook-2013}.
Thus it is enough to derive a closed-form expression of the cross-entropy between two Cauchy densities to get the closed-form formula for the KL divergence between these densities.
Observe that in general, the cross-entropy $-\int p(x) \ln q(x) \dmu(x) $ between two densities $p$ and~$q$ may not be finite when the integral diverges.
The cross-entropy and entropy for densities belonging to an exponential family have been reported in~\cite{HEF-2010}.

\section{Closed-form formula}

We have
\begin{eqnarray}
h^\times(p_{l_1,s_1}:p_{l_2,s_2}) &=& -\int \frac{s_1}{\pi (s_1^2+(x-l_1)^2)} \log \frac{s_2}{\pi (s^2_2+(x-l_2)^2)} \dmu(x),\\
&=& \log\frac{\pi}{s_2} + \frac{s_1}{\pi} \int \frac{\log (x^2-2l_2x+s_2^2+l_2^2) }{x^2-2l_1x+s_1^2+l_1^2}\dmu(x),\\
&=& \log\frac{\pi}{s_2} + \frac{s_1}{\pi} A(1,-2l_1,l_1^2+s_1^2 ; 1,-2l_2,l_2^2+s_2^2), \label{eq:h_cross}
\end{eqnarray}
where we define the parametric definite integral $A$ as follows:
\begin{equation}
A(a,b,c;d,e,f) = \int_{-\infty}^\infty \frac{\log(dx^2+ex+f)}{ax^2+bx+c} \dx,
\end{equation}
for both positive polynomials $ax^2+bx+c$ and $dx^2+ex+f$ on the reals.
(For a positive polynomial $ux^2+vx+w$ on the reals, we mean a polynomial given by coefficients satisfying $u > 0$, $v \in \bbR$, $w > 0$, and $4uw-v^2 > 0$.)
In particular, the polynomials $x^2-2l_1x+s_1^2+l_1^2$ and $x^2-2l_2x+s_2^2+l_2^2$ are both positive.

Using the {\sc Mgfun} package\footnote{\url{https://specfun.inria.fr/chyzak/mgfun.html}} for the computer algebra system Maple, we calculate the definite integral $A$ (see Appendix~\ref{sec:maple} for details), and get

\begin{proposition}
We have
\begin{equation} \label{eq:A}
A(a,b,c;d,e,f) = \frac{2\pi\left(\log(2af-be+2cd+\sqrt{4ac-b^2}\sqrt{4df-e^2})-\log (2a)\right)}{\sqrt{4ac-b^2}}.
\end{equation}
\end{proposition}

Thus we get overall the following expression for the Kullback-Leibler divergence between Cauchy densities (see the end of Appendix~\ref{sec:maple}):

\begin{theorem}
The Kullback-Leibler divergence between Cauchy density $p_{l_1,s_1}$ and $p_{l_2,s_2}$ is
\begin{equation}
\KL(p_{l_1,s_1}:p_{l_2,s_2})= \log\frac{(s_1+s_2)^2+(l_1-l_2)^2}{4s_1s_2}. \label{eq:KLCauchy}
\end{equation}
\end{theorem}

Observe that the KL divergence is always finite and symmetric:
\begin{equation}
\KL(p_{l_1,s_1}:p_{l_2,s_2})=\KL(p_{l_2,s_2}:p_{l_1,s_1}).
\end{equation}

The symmetric property does not hold in general for location-scale families (e.g., normal family).

In particular, when $l_1=l_2=l$ (scale family), we have
\begin{equation}
\KL(p_{l,s_1}:p_{l,s_2})=2\log\frac{s_1+s_2}{2\sqrt{s_1s_2}},
\end{equation}
recovering the result stated in~\cite{JSMeans-2019}.

When $s_1=s_2=s$ (location family), we have
\begin{equation}
\KL(p_{l_1,s}:p_{l_2,s})= \log \left(1+\frac{(l_1-l_2)^2}{4s^2}\right).
\end{equation}

The cross-entropy between two location-scale Cauchy distributions is
\begin{equation}
h^\times(p_{l_1,s_1}:p_{l_2,s_2})= \log \frac{\pi((s_1+s_2)^2+(l_1-l_2)^2)}{s_2}.  \label{eq:hcrossCauchy}
\end{equation}
It follows that the differential entropy is $h(p_{l,s})=h^\times(p_{l,s}:p_{l,s})=\log 4\pi s$, in accordance with~\cite{h-handbook-2013}.

By the same change of variable in the KL definition Eq.~\ref{eq:KLls} as the one used in the appendix, we find the following key property~\cite{KLls-2019} of the KL divergence between location-scale densities:
\begin{equation}
\KL(p_{l_1,s_1}:p_{l_2,s_2}) = \KL(p_{0,1}:p_{\frac{l_2-l_1}{s_1},\frac{s_2}{s_1}}).
\end{equation}
Note that this property holds for any $f$-divergence in general, that is for integrals
\[ \int f\left(\frac{p(x)}{q(x)}\right) q(x) \dmu(x) \]
given for two densities by a fixed function~$f$.

Indeed, we check that
\begin{eqnarray}
 \KL(p_{0,1}:p_{\frac{l_2-l_1}{s_1},\frac{s_2}{s_1}}) &=& \log \frac{\left(1+\frac{s_2}{s_1}\right)^2 +\left(0-\frac{l_2-l_1}{s_1}\right)^2}{4\frac{s_2}{s_1}},\\
&=&  \log\frac{(s_1+s_2)^2+(l_1-l_2)^2}{4s_1s_2},\\
&=& \KL(p_{l_1,s_1}:p_{l_2,s_2}).
\end{eqnarray}

\vskip 0.5cm

\noindent {\bf Acknowledgments.} Frank Nielsen would like to thank Joris van der Hoeven (Ecole Polytechnique, LIX) for mentioning the {\sc Mgfun}  package of Frédéric Chyzak.

\appendix

\section{A Maple-assisted calculation of the definite integral~$A$ and of the KL divergence}\label{sec:maple}

To the best of our knowledge,
the closed-form formula for the definite integral~$A$
is only given with no explicit proof in the literature,
in an almost equivalent form \cite[Eq.~19, page 514]{PrudnikovBrychkovMarichev-1986-IS1}
as
$$
\int_{-\infty}^\infty \log(a^2-2abx+x^2) \frac{\dx}{x^2+z^2} = \frac{\pi}{z}\log(z^2+2az\sqrt{1-b^2}+a^2),\quad
a>0, \ \Re z>0, \ b\in (-1,1].
$$
For real~$z$, this integral is a specialization of our more general integral $A$.
The implicit proof there is,
following an algorithm by Marichev described in \cite{AdamchikMarichev-1990-ACI},
to express the integrand as a product of so-called Meijer G-functions,
to use one of the formulas in \cite[Sec. 2.24]{PrudnikovBrychkovMarichev-1990-IS3},
providing a ``closed-form'' formula as another Meijer G-function,
to identify this formula as the right-hand side of~\eqref{eq:A}.

For an explicit proof, we prefer another approach,
named creative telescoping,
that makes algorithmic use of derivation under the integral sign.
The algorithm that we use can be viewed as an extension \cite{Chyzak-2000-EZF}
applicable to integrands satisfying higher-order linear differential equations
of an earlier algorithm for hyperexponential integrands \cite{AlmkvistZeilberger-1990-MDI},
that is, satisfying first-order equations.
(See also \cite{Chyzak-2014-ACT} for an introduction to developments of creative telescoping.)

The calculation starts with proving the special case $(a, b, c) = (1, 0, 1)$.
The integrand is then
\[ \phi(d, e, f; x) = \frac{\ln(dx^2+ex+f)}{x^2+1} . \]
It satisfies second-order homogeneous linear differential equations,
one for each of the variables $d$, $e$, $f$, $x$,
as well as relations between the first-order derivatives.
Chyzak's implementation of~\cite{Chyzak-2000-EZF} as the {\sc Mgfun} package for Maple readily derives
the nonhomogeneous relation
\begin{equation}\label{eq:ct}
\calL\frac{\der\phi}{\der d}(d,e,f; x) = \frac{\der\psi}{\der x}(d,e,f; x) ,
\end{equation}
where we define a linear differential operator~$\calL$ by
\begin{multline*}
\calL y =
(2df+e^2+6f^2)(4df-e^2)(d^2-2df+e^2+f^2) \frac{\der^3y}{\der d^3} \\
+(60d^3f^2+24d^2e^2f+132d^2f^3-6de^4-60de^2f^2-252df^4+18e^4f+108e^2f^3+60f^5) \frac{\der^2y}{\der d^2} \\
+(96d^2f^2+60de^2f+336df^3-6e^4-84e^2f^2-240f^4) \frac{\der y}{\der d}
+24(df+e^2+5f^2)f y ,
\end{multline*}
as well as
\begin{equation*}
\psi(d,e,f; x) = \frac{-2x P(d,e,f; x)}{(dx^2+ex+f)^2} \frac{\der\phi}{\der d}(d,e,f; x)
\end{equation*}
for the polynomial
\begin{multline*}
P(d,e,f; x) = e(d^2f^2-4de^2f-9df^3-e^4-7e^2f^2-6f^4)x^5 \\
-3f(3de^2f+4df^3+2e^4+11e^2f^2+4f^4)x^4 +e(d^2f^2-4de^2f-18df^3-e^4-16e^2f^2-51f^4)x^3 \\
-f(9de^2f+16df^3+6e^4+37e^2f^2+32f^4)x^2 -9ef^2(df+e^2+5f^2)x -4f^3(df+e^2+5f^2) .
\end{multline*}
Observe that \eqref{eq:ct}~is really a differential relation on the rational function
\[ \frac{\der\phi}{\der d}(d,e,f; x) = \frac{x^2}{(dx^2+ex+f)(x^2+1)} , \]
which is indefinitely differentiable with respect to~$x$ on the integration interval~$\bbR$,
and that
\begin{equation*}
\lim_{x\rightarrow+\infty} \psi(d,e,f; x) = \lim_{x\rightarrow-\infty} \psi(d,e,f; x) = -2e(d^2f^2-4de^2f-9df^3-e^4-7e^2f^2-6f^4)/d^3 .
\end{equation*}
As a consequence, \eqref{eq:ct}~can be integrated over~$\bbR$,
leading to
\[ \calL \frac{\der A}{\der d}(1,0,1; d,e,f) = 0 . \]
The generation and analysis leading to the previous homogeneous differential equation is done in Maple as follows.
\begin{quote}
{\footnotesize
\begin{verbatim}
libname := "./Mgfun.mla", libname:
assms := a>0, b :: real, c>0, 4*a*c-b^2 > 0, d>0, e :: real, f>0, 4*d*f-e^2 > 0;
phi := ln(d*x^2+e*x+f)/(x^2+1);
ct := Mgfun:-creative_telescoping(phi, d::diff, x::diff)[1];
L := collect(subs(diff(_F(d), d) = y(d), ct[1]), diff, factor);
map(p -> collect(p, x, factor), ct[2]);
P := collect(ct[2] / (-2*x*diff(_f(d,x),d)) * (d*x^2+e*x+f)^2, x, factor);
diff(phi, d);
psi := factor(eval(ct[2], _f(d,x) = phi));
limit(psi, x=+infinity);
limit(psi, x=-infinity);
%% - %;
\end{verbatim}}
\end{quote}

Now, note that the Cauchy--Lipshitz theorem applies to the differential equation $\calL y = 0$ at~$d = 0$.
Maple can indeed find algebraic expressions
for the integrals
\[ \frac{\der^i A}{\der d^i}(1,0,1; d,e,f) = \int_{-\infty}^\infty \frac{\der^i\phi}{\der d^i}(d,e,f; x) \dx \]
when~$1 \leq i \leq 4$.
Such expressions are in $\bbQ(\pi,d,e,f,\sqrt{4df-e^2})$.
Maple can then solves the differential equation $\calL y = 0$
under the constraints $\frac{\der^i y}{\der d^i}(d,e,f) = \frac{\der^i A}{\der d^i}(1,0,1; d,e,f)$, thus proving
\begin{equation}
\frac{\der A}{\der d}(1,0,1; d,e,f) = \pi \frac{(d-f)(4df-e^2)+(-2df+e^2+2f^2)\sqrt{4df-e^2}}{(d^2-2df+e^2+f^2)(4df-e^2)} .
\end{equation}
Maple can then compute a primitive with respect to~$d$,
which provides~$A(1,0,1; d,e,f)$
\begin{equation}
A(1,0,1; d,e,f) = \frac\pi2(\ln{G_1}-\ln{G_2}+\ln{G_3}) + K(e,f)
\end{equation}
where
\begin{align}
G_1(d,e,f) &= d+f+\sqrt{4df-e^2} , \\
G_2(d,e,f) &= d+f-\sqrt{4df-e^2} , \\
G_3(d,e,f) &= d^2-2df+e^2+f^2 = (d-f)^2 + e^2 = G_1(d,e,f) G_2(d,e,f)
\end{align}
and $K$~is a function~$K$ still to be determined.
\begin{quote}
{\footnotesize
\begin{verbatim}
evala(Normal(dsolve({
  eval(ct[1], diff(_F(d), d) = y(d)),
  seq(diff(y(d), [d$i]) =
        evala(Normal(int(diff(phi, [d$(i+1)]), x=-infinity..infinity)
          assuming assms)), i=0..3)
})));
A := simplify(int(op([1,2], %), d)) assuming assms;
\end{verbatim}}
\end{quote}
Note that an alternative, more direct derivation up to this point is by observing
that $\phi(d, e, f; x)$ and~$\frac{\der\phi}{\der d}(d, e, f; x)$ are both integrable on~$\bbR$,
so that one has
\[ A(1,0,1; d,e,f) = \int \left( \int_{-\infty}^\infty \frac{\der\phi}{\der d}(d, e, f; x) \dx \right) \dt .
\]
In Maple, this leads to the same formula as with the method of creative telescoping, using the following commands:
\begin{quote}
{\footnotesize
\begin{verbatim}
assms := a>0, b :: real, c>0, 4*a*c-b^2 > 0, d>0, e :: real, f>0, 4*d*f-e^2 > 0;
phi := ln(d*x^2+e*x+f)/(x^2+1);
A := simplify(int(int(diff(phi, d), x=-infinity..infinity), d)) assuming assms;
\end{verbatim}}
\end{quote}
In a first step, the previous, one-line derivation calculates a primitive with respect to~$x$:
\begin{multline*}
B(d,e,f; x) = \int_0^x \frac{\der\phi}{\der d}(d, e, f; y) \dy = \\
\frac{
2 \sqrt{4df-e^2}\left((d-f)\arctan x - \frac e2\ln(dx^2+ex+f) + \frac e2\ln(x^2+1)\right)
- 4 (df-\frac{e^2}2-f^2)\arctan\frac{2dx+e}{\sqrt{4df-e^2}}
}{(2d^2-4df+2e^2+2f^2)\sqrt{4df-e^2}} .
\end{multline*}
Then, the integral~$A$ can be derived as a difference of limits of~$B$ at $x = \pm\infty$.
For the primitivation, Maple uses the Risch algorithm for determining if a primitive of a so-called Liouvillian function can be expressed as a Liouvillian function \cite{Risch-1969-PIF,Risch-1970-SPI}.
This includes the case of rational expressions in logarithms, like~$\phi$,
and for~$\phi$ a primitive is in the class.
The primitive~$B$ can be obtained in Maple by the following command:
\begin{quote}
{\footnotesize
\begin{verbatim}
B := simplify(int(diff(phi, d), x)) assuming assms;
\end{verbatim}}
\end{quote}

As $G_1$ and~$G_3$ are positive by construction when~$d > e^2/4/f$,
so is~$G_2$, and thus, $\ln G_3 = \ln G_1 + \ln G_2$, so that
$A(1,0,1; d,e,f) = \pi \ln G_1 + K(e,f)$.
To fix the integration constant~$K$, we successively take derivatives with respect to $f$ and~$e$, and use the fact that Maple is, like above for $\frac{\der\phi}{\der d}$, able to integrate $\frac{\der\phi}{\der f}$ and~$\frac{\der\phi}{\der e}$ with respect to~$x$.
This proves that $K$~is in fact independent of $f$ and~$e$.
Choosing next $f = d$ and~$e = 2d$ proves~$K = 0$,
and the final formula for the special case $(a,b,c) = (1,0,1)$ is:
\begin{equation}
A(1,0,1; d,e,f) = \pi \ln(d+f+\sqrt{4df-e^2}) .
\end{equation}
\begin{quote}
{\footnotesize
\begin{verbatim}
(d+f+sqrt(4*d*f-e^2)) * (d+f-sqrt(4*d*f-e^2));
A := simplify(subs(expand(%) = %, A)) assuming op(1, %) > 0, op(2, %) > 0;
simplify(int(diff(phi, f), x=-infinity..infinity) - diff(A, f)) assuming assms;
simplify(int(diff(phi, e), x=-infinity..infinity) - diff(A, e)) assuming assms;
{f = d, e = 2*d};
simplify(int(eval(phi, %), x=-infinity..infinity) - eval(A, %)) assuming assms;
EQN := Int(phi, x=-infinity..infinity) = A;
\end{verbatim}}
\end{quote}

Dealing with the general case next amounts to reducing it to the previous formula,
by a suitable change of integration parameter.
Namely, after setting
\begin{equation}
D = \frac{d(4ac-b^2)}{4a^2} , \quad
E = \frac{(ae-bd)\sqrt{4ac-b^2}}{2a^2} , \quad
F = \frac{4a^2f-2abe+b^2d}{4a^2} , \quad
K = \frac{2}{\sqrt{4ac-b^2}} ,
\end{equation}
one has
\[ A(a,b,c; d,e,f) = K A(1,0,1; D,E,F) . \]
This is done in Maple as follows:
\begin{quote}
{\footnotesize
\begin{verbatim}
phi := ln(d*x^2+e*x+f)/(a*x^2+b*x+c);
alt_phi := eval(factor(normal(
  subs(x = (x*sqrt(4*a*c-b^2)-b)/2/a, phi) * sqrt(4*a*c-b^2)/2/a)),
  ln = (e -> ln(collect(e, x, normal))));
K := eval(alt_phi, ln=1)*(x^2+1);
alt_phi := alt_phi / K;
op(select(has, alt_phi, ln));
local D := coeff(%, x, 2);
E := coeff(%%, x, 1);
F := coeff(%%%, x, 0);
GEN_EQN := Int(phi, x=-infinity..infinity) =
  K * simplify(eval(op(2, EQN), {d = D, e = E, f = F})) assuming assms;
\end{verbatim}}
\end{quote}

The cross-entropy, then our expression~\eqref{eq:KLCauchy} for the KL divergence,
follow by appealing to the same kind of simplification as above
in, respectively, \eqref{eq:hcrossCauchy} and \eqref{eq:KL-from-h}:
\begin{quote}
{\footnotesize
\begin{verbatim}
GEN_A := op(2, GEN_EQN);
h_cross := simplify(log(Pi/s2) + s1/Pi * eval(GEN_A,
  {a=1, b=-2*l1, c=l1^2+s1^2, d=1, e=-2*l2, f=l2^2+s2^2})) assuming s1>0, s2>0;
KL := simplify(h_cross - eval(h_cross, {l2=l1, s2=s1})) assuming s1>0, s2>0;
\end{verbatim}}
\end{quote}

As a final note, we remark that we have used the Maple syntax \verb+... assuming assms+
to allow the automatic use of analytic formulas for logarithms and square roots,
both in the procedures \verb+ln+ and \verb+sqrt+ for mathematical functions,
and in the procedures \verb+int+ and \verb+simplify+ for integration and simplification.
We have found this more efficient than using the \verb+assume(...)+ facility,
which would have attached properties to the names \verb+a+, \dots, \verb+f+, \verb+x+.


\end{document}